# Earth as an Exoplanet: A Two-dimensional Alien Map

Siteng Fan[1], Cheng Li[1], Jia-Zheng Li[1], Stuart Bartlett[1],

Jonathan H. Jiang[2], Vijay Natraj[2], David Crisp[2], Yuk L. Yung[1,2]

1: Division of Geological and Planetary Sciences, California Institute of Technology, Pasadena, CA 91125, USA

2: Jet Propulsion Laboratory, California Institute of Technology, Pasadena, CA 91109, USA

Corresponding author: Siteng Fan (stfan@gps.caltech.edu)

## Abstract

Resolving spatially-varying exoplanet features from single-point light curves is essential for determining whether Earth-like worlds harbor geological features and/or climate systems that influence habitability. To evaluate the feasibility and requirements of this spatial feature resolving problem, we present an analysis of multi-wavelength single-point light curves of Earth, where it plays the role of a proxy exoplanet. Here, ~10,000 DSCOVR/EPIC frames collected over a two-year period were integrated over the Earth's disk to yield a spectrally-dependent point source and analyzed using singular value decomposition. We found that, between the two dominant principal components (PCs), the second PC contains surface-related features of the planet, while the first PC mainly includes cloud information. We present the first two-dimensional (2D) surface map of Earth reconstructed from light curve observations without any assumptions of its spectral properties. This study serves as a baseline for reconstructing the surface features of Earth-like exoplanets in the future.

## 1. Introduction

Since the first exoplanet was detected (Campbell et al. 1988), approximately four thousand more have been confirmed (NASA Exoplanet Archive, exoplanetarchive.ipac.caltech.edu). Among these exoplanets, a number of them have similar properties to Earth, and may be habitable, e.g. TRAPPIST-1e (Gillon et al. 2017). However, existing measurements are still not adequate to determine whether these planets can support life. A geological and climate system that supports all three phases of water is critical to life on Earth's surface. The presence of atmospheric water vapor, clouds and surface oceans could therefore serve as biosignatures that can be observed from a distance, and are also among the indicators for habitability. Identifying surface features and clouds on exoplanets is thus essential in this context.

Earth is the only known planet that harbors life. Remote sensing observations of Earth can therefore serve as proxies for a habitable exoplanet, as seen from the perspective of hypothetical distant observers. A number of such studies have been performed since the first analysis of snapshots of Earth obtained by the Galileo spacecraft (Sagan et al. 1993, Geissler et al. 1995). Two observations of Earth's light curve, each spanning one day, obtained by the Deep Impact spacecraft, were used to identify changes of surface features (land/ocean) and clouds (Cowan et al. 2009, 2011; Cowan & Strait 2013). Using principal component analysis (PCA), time series of disk-integrated spectra of Earth were decomposed into two dominant "eigencolors", which contained 98% of the light curve variance. Land/ocean changes and cloud patterns were recently extracted from Earth's light curves using two years of observations of the Earth's bright side from the Deep Space Climate Observatory (DSCOVR, Jiang et al. 2018). Besides the Earth, other solar system planets (e.g. Jupiter, Ge et al. 2019) have also been treated as proxy exoplanets, with their light curves analyzed to provide baselines for exoplanet studies.

Despite interference from clouds, two-dimensional (2D) surface maps of exoplanet surfaces can be constructed using time-resolved spectra together with orbital and viewing geometry information, which in principle can be derived from light curves and other observables. Spatial maps of hot Jupiter atmospheres have been constructed (e.g. Knutson et al. 2007, Louden & Wheatley 2015). However, the detection and mapping of potentially habitable Earth-like exoplanets remains a challenge, especially when surface features and clouds contribute strongly to light curves. Numerical models that simulate light curve observations of the Earth using known spectra of different surface types have been used to test methods for retrieving 2D maps (e.g. Fujii & Kawahara 2012, Cowan & Fujii 2017, Farr et al. 2018). Using two-day single-point light curves from NASA's EPOXI mission, Cowan et al. (2009) presented the first retrieved longitudinal surface map of Earth's surface. However, 2D surface maps have not yet been derived from actual single-point light curve observations, due to their low temporal or spatial resolutions.

In this paper, we reanalyze the two-year DSCOVR/EPIC observations presented by Jiang et al. (2018) to study the Earth as a proxy exoplanet. We integrate over the disk of the Earth to reduce each image to a single point source, in order to simulate the light curve of a distant exoplanet. We report the first 2D surface map of this proxy cloudy exoplanet reconstructed from its single-point light curves, without making any assumptions about its spectral features.

## 2. Observations and labels

Disk-integrated light curves analyzed in this work are derived from Earth's images obtained by DSCOVR's Earth Polychromatic Imaging Camera (EPIC; www.nesdis.noaa.gov/DSCOVR/spacecraft.html) during 2016 and 2017. The DSCOVR spacecraft is positioned at the first Sun-Earth Lagrangian point (L1), viewing the sunlit face of Earth from a distance of about $10^6$ km. From this vantage point, DSCOVR views the entire disk of Earth, illuminated near local noon. This provides an ideal geometry for studying the Earth as a proxy exoplanet seen near opposition relative to its parent star. Since the observations are all near full-phase configuration, similar to those near secondary eclipse (when the planet is blocked by the star), the phase-angle effect (Jiang et al. 2018) is not considered in this work. The EPIC instrument images the Earth every 68–110 minutes, returning a total of 9740 frames over a two-year period (2016–2017), with a 2048×2048 charge-coupled device (CCD) in 10

narrowband channels (317, 325, 340, 388, 443, 552, 680, 688, 764 and 779nm), which are selected primarily for investigations of the Earth's climate. A sample observation of reflectance in the 680nm channel at 9:27 UTC, 2017 February 8[th] is shown in Figure 1a.

We integrate the spatially-resolved images over the Earth's disk to simulate observations of an exoplanet that is detected as a point source. This results in a mean reflectance of 0.22 for the image shown in Figure 1a. At each time step, reflectance images obtained sequentially from the 10 channels, with exposure time ranging from 22 ms to 654 ms, are combined to form a 10-point reflection spectrum. Since the surface materials and cloud distributions on an Earth-like exoplanet may be significantly different from those on Earth, we do not assume any known spectral features of the surface; instead, we label each reflection spectrum with disk-averaged fractions of land and clouds on the sunlit face of the Earth to evaluate the cause of the change in Earth's light curves. The two fractions are computed as weighted averages, where the weights are proportional to the cosine of the solar zenith angle. Figure 1b shows a land/ocean map as seen from the same viewing angle, with a land fraction of 0.33 as defined by the Global Self-consistent, Hierarchical, High-resolution Geography Database (Wessel et al. 1996). We use the Level-3 MODIS Atmosphere Daily Global Product (Platnick et al. 2015) to compute the cloud fraction label at each time step. Cloud fractions are linearly interpolated between days, since there are multiple EPIC observations each day. This results in a 0.61 cloud fraction for the example observation shown in Figure 1c.

In summary, using EPIC observations collected during 2016 and 2017 and the concomitant viewing geometry, we obtain a time series consisting of 10-point reflection spectra with two coverage fraction labels, which are combined in Section 3 to analyze the time series of light curves and to recover the surface map of the proxy exoplanet. The mean and variance of the reflectance in each channel are shown in Figure 1d; clearly, there is considerable difference among the channels. In the following analysis, we normalize the time series of reflection spectra to yield zero means and unit standard deviations, since the singular value decomposition (SVD), presented in Section 3, is sensitive to the scaling between dimensions of the dataset. This is done in order to give each channel equal importance, since the wavelengths are likely to be different for future exoplanet observations.

## 3. Time series analysis

Time series of disk-averaged light signals carry information about the spectral variability of exoplanets. Analyzing such signals from Earth (in this case, serving as a proxy exoplanet) could provide a baseline for future exoplanet studies. Jiang et al. (2018) used the same dataset and analyzed irradiances from individual EPIC channels to correlate their changes with different types of reflective surfaces, using the original high-resolution images and known spectra of materials on Earth. In their analysis, Jiang et al. (2018) qualitatively explained the variations of single-point light curves for the years 2016 and 2017. However, for future exoplanet studies, Earth-like planets will only be resolved as disk-integrated point sources. Therefore, methods are needed to retrieve information about exoplanet environments from these disk-integrated, point-source observations. For Earth, the light curve is dominated primarily by the cloud cover and land/ocean fraction. Here, we adopt these two parameters, viz., the land and cloud fraction labels, as metrics for evaluating the success of our analysis technique. Therefore, all types of land surfaces are considered the same and an "averaged" land is used in the analysis. Spectra of any reflective surfaces are assumed to be unknown since those on exoplanets could be very different.

We use SVD to decompose the time series into principal components (PCs), and then separate the influences of different reflective surfaces. The first two PCs of the scaled light curves, or "eigencolors" as defined by Cowan et al. (2009), are shown in Figure 1e. Given that the singular values are the square roots of the variance along corresponding dimensions, the first two PCs contain 96.2% of the variation of the scaled light curves (Figure 2a). The ordering of the variance accounted for by the PCs depends on the contribution of the land and cloud fractions to the time series variance. Although the PCs are orthogonal, changes in land and cloud fraction are not independent from each other. Land and ocean are fixed to Earth's surface, and appear periodically in the time series of the light curve. Clouds are variable; some parts of Earth are perennially covered by clouds (e.g., the southern oceans), while some others are always clear (e.g., the Sahara Desert). Therefore, some clouds may be synchronized with the rotation of Earth's surface into and out of the instrument field of view, precluding the separation of these clouds from surface features based solely on single-point observations. Nevertheless, this information is encoded in the time series of the disk-integrated image. Information about the perennially cloud or perennially clear scenarios (hereafter referred to as surface-correlated clouds) is expected to be included in the same PC for land/ocean, while that of the rest of clouds would be in another PC.

In order to interpret the physical meaning of the PCs, it is necessary to analyze the relationship between the two labels (land and cloud fractions) and the reflectance time series. We use a machine learning method, called Gradient Boosted Regression Trees (GBRT; Friedman 2001), to evaluate the importance of the PCs for each label (Figure 2a). This technique computes the relative importance of each PC and denotes them with weights, where the sum of the weights is normalized to unity. Further details on implementing the GBRT technique are presented in Appendix a. Clearly, the first two PCs (hereafter, PC1 and PC2) are the most critical, while the contributions from other PCs are negligible. They also have considerably different weights for the land and cloud fractions. For the land fraction, the weight of PC2 is 0.88, while that of PC1 is only 0.03 (Figure 2a). Changes in the land/ocean fraction are independent of PC1 and mostly correlated with PC2 ($r^2$=0.91; Figure 2b). For the cloud fraction, the two PCs have comparable weights, 0.28 for PC1 and 0.49 for PC2 (Figure 2a). Given the strong correlation between PC2 and surface features, the comparable importance of PC1 and PC2 for clouds suggests that the clouds consist of two types: surface-independent clouds and surface-correlated clouds. This confirms the conclusions of our qualitative analysis above that some changes in clouds can correlate temporally with the surfaces underneath. The importance of clouds in PC2 is likely to be due to these surface-correlated clouds. Conversely, PC1 contributes the largest variation to the light curves via surface-independent clouds that are not correlated with the land/ocean fraction.

The interpretation of the first two PCs is also supported by their time series (Figure 3). The time series for PC1 (Figure 3a) has more scatter than that for PC2 (Figure 3c). The envelope of daily maximum and minimum for PC2 changes gradually between consecutive days, while that for PC1 changes drastically, which indicates that PC2 is more likely to represent features that are surface-associated than detached. Moreover, PC2 peaks in summer, when the northern hemisphere is facing the sun and the spacecraft. The change of PC2 within a day is constant throughout most of the observations, due to the diurnal cycle and the Earth's longitudinal land fraction asymmetry. Fourier analysis shows that both PC1 and PC2 have annual, semi-annual, diurnal and half-daily cycles (Figures 3b and 3d). The diurnal cycle of PC2 has the strongest signal in the power spectrum, while that of PC1 is relatively weak. This may be due to the fact that land/ocean reappear with small changes between two consecutive days, while surface-independent clouds can be significantly different in the same time period.

Although convoluted, information on the spatial distribution of different types of surfaces and clouds is fully contained in the time series of an observed planet's light curves. As discussed above, we separate the clouds from surface features using SVD. Surface information about the Earth is mostly contained in PC2 with a strong linear correlation. Here we report the first 2D surface map of Earth (Figure 4a) that is reconstructed from single-point light curves using the following assumptions. For the purpose of retrieving the map, the viewing geometry is assumed to be known in this work and obtained from DSCOVR navigation data based on maneuvers that took place during the two-year observation period. It can, in principle, also be derived using light curves and other data (e.g., radial velocity, transit timing) as discussed in more detail in Section 4. In the construction of Earth's surface map, spectral features of reflective surfaces are assumed to be unknown in order to facilitate generalization for future Earth-like exoplanet observations. We make the minimal assumptions that the incoming solar flux is uniform and known, and that the entire surface of the proxy exoplanet acts as a Lambertian reflector. Although Earth's ocean is strongly non-Lambertian, we still employ the Lambertian assumption because we assume that the surface properties are unknown. This may overestimate the ocean contribution at large distances from the specular point (glint spot) and underestimate it at the specular point. With these assumptions, constructing the Earth's surface map becomes a linear regression problem. Mathematical details are provided in Appendix b, and uncertainty estimation is discussed in Appendix c. We set the regularization parameter to be $10^{-3}$ for producing the optimal surface map; results for other values are given in Appendix b. The quantity derived in the map is the value of PC2, which has a positive linear correlation with the land fraction as noted above. Coastlines in the reconstructed map are determined by the median value of PC2, which is consistent with the minimal assumption of the overall land fraction being unknown. Compared with the true land/ocean map (Figure 4b), the retrieved map successfully recovers all of the major continents, while there exist some disagreements over oceans. This may be due to the fact that there is often significant cloud coverage over oceans, which reduces sensitivity to surface information in the observations.

## 4. Discussion

A critical requirement for constructing a 2D surface map for an exoplanet is the assumption that the surface information can be extracted from light curves. In the case of Earth, acting as a proxy exoplanet, SVD of light curves can successfully separate Earth's surface from surface-independent clouds. However, the relationships between PCs and features may not be the same for Earth-like exoplanets. This will depend on whether the surface type or clouds introduce measurable variations in the light curves. For Earth, the surface-independent clouds contribute 70.3% of the total variance of scaled light curves; the contribution of the surface is 25.9% (Figure 2a). On an exoplanet with less cloud cover, the ratio of these two numbers can be different or even less than unity, which would result in a switch between the first two PCs. In extreme cases where an exoplanet is either cloudless or fully covered by clouds, there will be only one dominant PC instead of the two comparable ones found in this analysis. The third or fourth PC may also be comparable if there exist changes that are significant in spatial and/or spectral scales, e.g. large-scale hydrological processes on continents or another layer of clouds with different composition. Since there will be no ground truth for an exoplanet, spectral analysis may distinguish the PCs between surface and clouds using appropriate assumptions about atmospheric and surface compositions. This can also be addressed by evaluating their time series. The PC associated with surface-independent features tends to have a more chaotic pattern in its time series (Figure 3a), while that associated with the surface is more likely to be periodic (Figure 3c). It is also worth noting that the surface features of an exoplanet, which are contained in one of the two PCs, will not necessarily be land and ocean. As long as two different surface types have a large albedo contrast and are non-uniformly distributed around the globe, one of the PCs would contain the changes. Materials that are detached from the surface, such as the surface-independent clouds in the case of the Earth, would also appear in one of the PCs if they have a large influence on the light curve. If these materials can be constrained by the reflection spectra, their fractions can be derived from the magnitude of the corresponding PC.

Once the surface information is extracted from light curves, the surface map of the exoplanet can be recovered from the observational geometry without making any spectral assumptions. Besides orbital elements, which can be determined from light curve observations, the only two geometry assumptions required for constructing the 2D surface map are the summer/winter solstice and the obliquity. The rotation period of a clear or partially-cloudy exoplanet can be inferred from the power spectrum of PCs using a Fourier transform (Figure 3d), which requires the observation frequency to be higher than that of the exoplanet's rotation. Studies have been performed to identify a planet's rotational period from light curves at different viewing geometries (e.g. Pallé et al. 2008). The summer/winter solstice would coincide with the maxima and minima of the PC time series as long as the asymmetry between the northern and southern hemispheres is noticeable, and the reflection changes monotonically with latitude when the sub-stellar point is near extremum. For the Earth, the peak of the time series of PC2 takes place on June 15[th], 2016 (Figure 3c), which is within one week of the true value. The obliquity of an exoplanet could be derived through its influence on the light curves. A number of recent publications (e.g. Schwartz et al. 2016; Kawahara 2016) developed methods for deriving the obliquity using its influence on the amplitude and frequency of light curves. Although these inversion methods are mostly based on a cloudless Earth and known surface spectral features, our SVD analysis of separating clouds from the surface could fill the gap.

Some issues exist in constructing and interpreting the retrieved Earth surface map. Degeneracy resulting from the convolution between pixel geometry and spectrum is the dominant factor affecting map construction quality. As discussed in Cowan & Strait (2013) and Fujii et al. (2017), light curves only cover a small portion of the PC plane, which results in a tradeoff between the spatial and spectral variation. Although only PC2 is used for constructing the surface map in this work, its time series covers only a small range of all valid values, whose corresponding land fractions are between 0 and 1. Therefore, we introduce a regularization parameter, $\lambda$, when constructing the map, to constrain pixel values of PC2 within a reasonable range; the resulting effect is described in Appendix b. We select the value of $\lambda$ based on the ground truth of Earth's surface map. Glints, features that are small in area but contribute significantly to the spectrum, may also influence the quality of the retrieved map. Their contributions to light curves are simulated and estimated in Lustig-Yaeger et al. (2018), and observational evidence in DSCOVR/EPIC images are reported by Li et al. (2018). In this work, the effect of glint is assumed to be on the "average ocean" and removed when scaling the light curves.

## 5. Summary

Spectrally-dependent, single-point light curves of the Earth were analyzed as observations of a proxy exoplanet. SVD analysis suggests that the majority of the information is captured by two principal components. The first captures the non-periodic behavior of surface-independent clouds. The second describes more periodic surface albedo structure. Using the fact that SVD separates the clouds from the surface, we derive the first 2D surface map of the Earth, acting as a proxy exoplanet, from single-point light curves, assuming only that the surface acts as a Lambertian reflector. The geometry is assumed to be known in the analysis, but in principle, it can be derived directly from light curves. This study serves as a baseline for analyzing observations of Earth-like exoplanets with unknown surfaces and possible clouds, enabling future assessments of habitability.

**Figures**

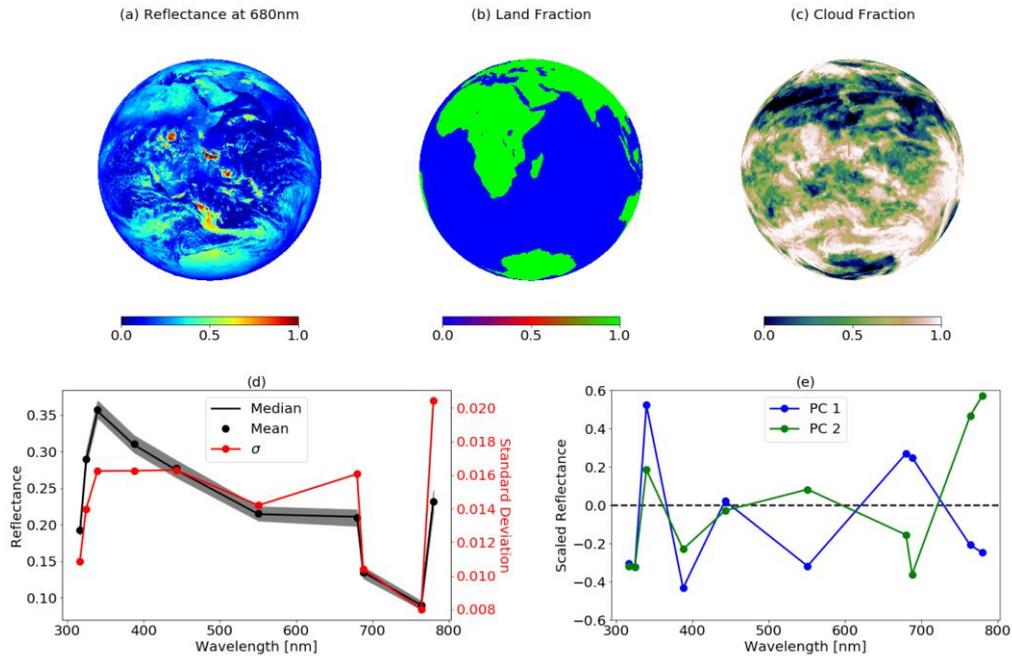

**Figure 1.** (a) Reflectance image in the 680nm channel of DSCOVR/EPIC obtained at 9:27 UTC, 2017 February 8[th]. The average reflectance is 0.22. (b) Land/ocean map of the Earth for the same scenario as (a), using the GSHHG database (Wessel et al. 1996). The average land fraction is 0.33. (c) Cloud fraction map for the same scenario as (a), obtained from the Level-3 MODIS Atmosphere Daily Global Product (Platnick et al. 2015). The averaged cloud fraction is 0.61. (d) Median (black solid line), mean (black dots) and standard deviation (red line and dots) of reflectance in each channel for ~10,000 DSCOVR/EPIC observations during the years 2016 and 2017. The grey shaded area shows the first and third quartiles of reflection spectra. (e) The first two principal components, PC1 (blue) and PC2 (green), of the scaled reflection spectrum time series.

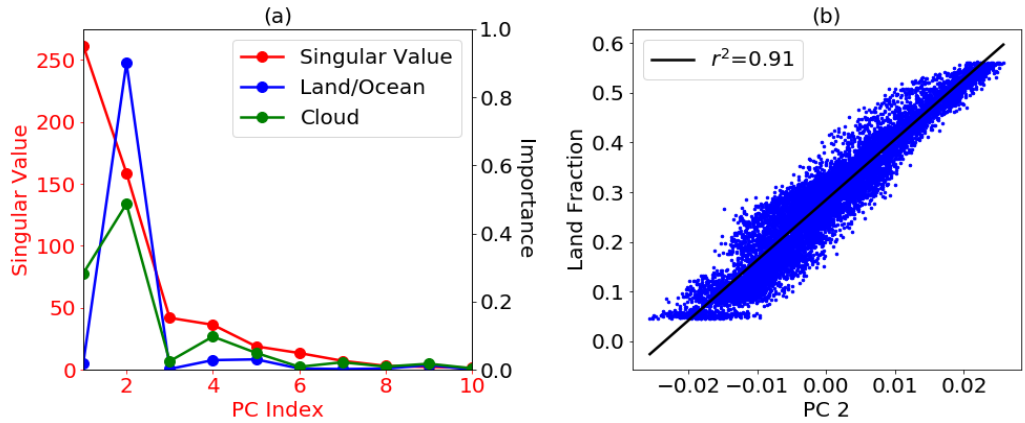

**Figure 2.** (a) Singular values of the principal components (PCs, red) and their importance to land (blue) and cloud (green) fractions. The importance of PCs for each fraction is evaluated using a Gradient Boosted Regression Trees model. (b) Scatter plot of the second principal component, PC2, as a function of land fractions (blue). The best fit line is shown in black, with a correlation coefficient of $r^2=0.91$.

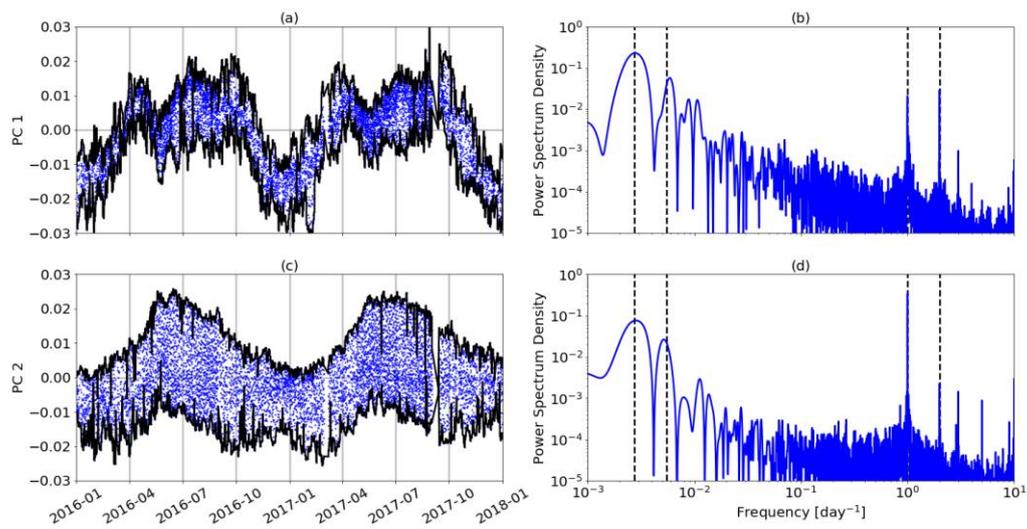

**Figure 3.** (a) Time series of the first principal component, PC1 (blue points). The envelops of daily maxima and minima are denoted by black lines. (b) Power spectrum of the time series of PC1. Cycles of annual, semiannual, diurnal and half-daily are denoted as black dashed lines. (c) and (d) are identical to (a) and (b), respectively, but correspond to PC2.

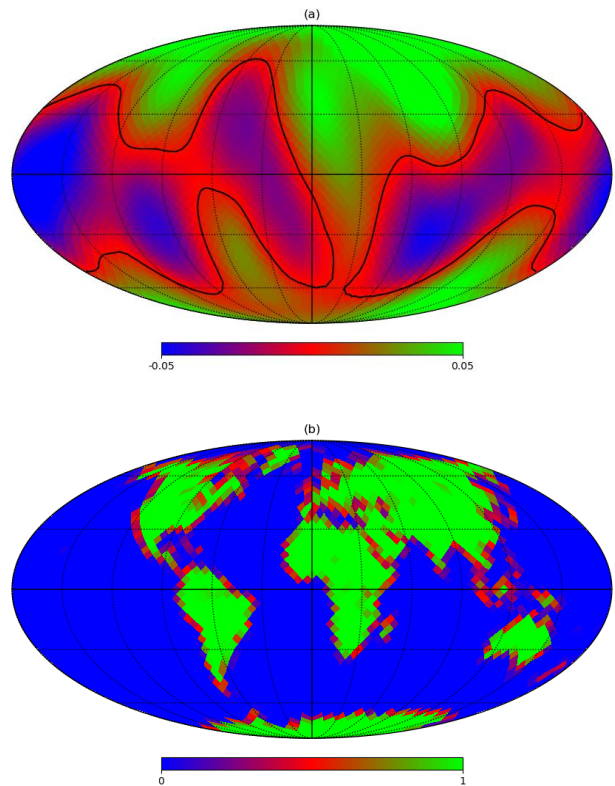

**Figure 4.** (a) 2D surface map of the Earth, treated as a proxy exoplanet, constructed using the PC2 time series. The contour of the median value is given by the black line, which serves as the coastline. The regularization parameter, $\lambda$, is $10^{-3}$ for constructing this map (see Appendix b for further details). (b) Global land/ocean map of the Earth.

**Acknowledgements**

This work was partly supported by the Jet Propulsion Laboratory, California Institute of Technology, under contract with NASA. YLY and DC acknowledge support by the Virtual Planetary Laboratory at the University of Washington. We thank Lixiang Gu and Mimi Gerstell for proofreading the manuscript.

**Appendix**

### a. Gradient Boosted Regression Trees

The decision tree model is a decision support tool, which is widely used in the field of machine learning. It uses a tree structure to classify data and make predictions. An example of decision trees used in this work is shown in Figure S1. At each parent node, data points are divided into two groups, called child nodes, by introducing a threshold for one of the PCs. The label of each child node, cloud fraction in this case, is computed as the mean value of data labels in this group. The mean squared error (MSE) between the prediction and individual labels in each node is evaluated accordingly. Given the finite number of probable ways to divide a parent node, there exists a best PC and a best threshold so that the average MSE of its two child nodes, weighted by their sample numbers, is minimized. Therefore, a decision tree can be constructed using this criterion from a root node, which contains all data points, and the labels are offset in order to have zero mean. For regularization, we introduce a maximum tree depth, a maximum total node number and a minimum node size to avoid over-fitting.

A machine learning technique, gradient boosting (Friedman 2001), is deployed to improve the model performance and reduce bias. Gradient boosting is an ensemble method that combines weak prediction models, shallow decision trees in this case, to make a final decision. The first decision tree is constructed using the original data points; starting from the second one, the decision tree fits the residual left by the previous tree. Therefore, the final decision is made as the sum of all decision trees. Since weak models tend to have large bias and small variance, while complex models have large variance and small bias, the boosting technique strikes a balance to achieve the minimum error. We use a Python package developed for machine learning, scikit-learn (Pedregosa et al. 2011), to develop this GBRT model. DSCOVR observations from the years 2016 and 2017 serve as training data, while those from the year 2018 are used as test data. The MSE of the test data is used to select the regularization parameters. The final GBRT model has 250 shallow decision trees, and each decision tree has a maximum depth of 5, a maximum total node number of 20 and a minimum node size of 100. One of the advantages of decision trees is that they can evaluate the feature importance (PCs in this work). We compute the importance of each PC as the number of times they appear as the threshold in the decision tree nodes, weighted by the number of node samples, and normalized to have a unit sum. Results of using land and cloud fractions as the labels are shown in Figure 2a. PC2 shows dominant correlation to the land fraction, while PC1 and PC2 show comparable correlation for clouds.

### b. Surface map construction

Given the assumptions discussed in Section 3, the averaged reflectance in the *i-th* channel, $R_i$, can be parameterized as follows:

$$R_i = \sum_p w_p r_{i,p} \quad (1)$$

where $r_{i,p}$ is the reflectance of the *p-th* pixel of an arbitrary map at the *i-th* wavelength; $w_p$ is the weight of the *p-th* pixel, which is determined by the viewing geometry and has the following form:

$$w_p = \begin{cases} c \cos(\alpha_p) \cos(\beta_p) & \text{when } \alpha_p < 90° \text{ and } \beta_p < 90° \\ 0 & \text{otherwise} \end{cases} \quad (2)$$

where $\alpha_p, \beta_p$ are the solar and the spacecraft zenith angles of the *p-th* pixel, respectively; $c$ is a normalization term such that the weights, $w_p$, sum to unity. The spacecraft is on a halo orbit around L1, which introduces differences between $\alpha_p$ and $\beta_p$. The sub-spacecraft point can have a solar zenith angle as large as ~7° at some time points. Due to the linearity of scaling and SVD, the averaged PC2 at each time point has the same form as the reflectance.

The observed time series of PC2, $v_t$ is given by:

$$v_t = \sum_p w_{t,p} x_p \quad (3)$$

where $x_p$ is the value of PC2 at the *p-th* pixel in the retrieved Earth map; $w_{t,p}$ has the same form as $w_p$ except that it varies with time. Including the entire time series, this becomes a linear regression problem:

$$\mathbf{W}_{[T \times P]} \mathbf{X}_{[P \times 1]} = \mathbf{V}_{[T \times 1]} \quad (4)$$

where $\mathbf{W}$, $\mathbf{X}$ and $\mathbf{V}$ are the matrices with elements $w_{t,p}$, $x_p$ and $v_t$, respectively. A penalty term, the $L^2$-norm of $\mathbf{X}$, is added to the squared error as a regularization parameter to prevent $x_p$ from deviating too far from zero and therefore avoid over-fitting. The regularized square error, $e$, can be expressed as:

$$e = |\mathbf{WX} - \mathbf{V}|_2^2 + \lambda |\mathbf{X}|_2^2 \quad (5)$$

where $\lambda$ is the regularization parameter. Consequently, the solution to minimizing e is as follows:

$$\mathbf{X} = (\mathbf{W^T W} + \lambda \mathbf{I})^{-1} \mathbf{W^T V} \quad (6)$$

where $\mathbf{I}$ is the identity matrix.

We use the Hierarchical Equal Area isoLatitude Pixelization method (HEALPix; Górski et al. 2005) to pixelate the retrieved map. This technique divides the Earth's surface into pixels with the same area and distributed uniformly on the sphere, appropriate for the DSCOVR observing geometry. The parameter $N_{side}$ in HEALPix is set to 16, which results in a 3072-pixel map with a spatial resolution of ~4°. After solving the regularized linear regression problem, we construct the first 2D surface map of Earth (Figure 4a) from single-point light curves.

The parameter $\lambda$ is selected using synthetic data, where elements in **V** are replaced by time series of the land fraction label. Three recovered maps using synthetic data are shown in Figure S2, where $\lambda$ has values of $10^{-4}$, $10^{-3}$ and $10^{-2}$, respectively. Comparing them with the ground truth, a value of $10^{-3}$ is seen to be optimal for $\lambda$. Due to the degeneracy in recovering maps from single-point observations and imperfect geometry assumptions, which includes unequal pixel weights and pixelization approximations, the map cannot be further improved even with perfect spectral observation. Fortunately, the map is not sensitive to $\lambda$ near its optimal value; changing it by one order of magnitude only results in small changes in the coastlines (Figures S2a and S2c). Therefore, we propose that $10^{-3}$ is a good choice for the regularization parameter for Earth-like exoplanets if observations have comparable numbers of pixels and time steps. The value should be adjusted according to the ratio of the two terms on the right hand side of Equation (5) when the number of pixels and/or time steps are different.

When the land fraction is not known, the selection of $\lambda$ becomes arbitrary, with the only constraint being that the resulting range of land fraction should be physically valid under the given assumptions. Two more possible maps reconstructed using time series of PC2 with different values of $\lambda$ ($10^{-4}$ and $10^{-2}$) are shown in Figure S3. Comparing the maps recovered using the observations and known ground truth (Figures 4a, S3 and S2), it is evident that clouds over oceans contribute considerably to the differences, since the land fraction label is not affected by clouds.

### c. Surface map uncertainty

We estimate the uncertainty in the retrieved Earth surface map (Figure 4a) in this section. The observation uncertainty is neglected, since at each time step ~$10^6$ pixels are averaged so that the observational uncertainty is reduced by a factor of ~$10^3$. Therefore, we mainly focus on the uncertainty in the linear regression presented in Appendix b.

We rewrite Equation (4) with a vector **U** for the "true values" of PC2 at each pixel as:

$$\mathbf{W}_{[T \times P]}\mathbf{U}_{[P \times 1]} + \boldsymbol{\varepsilon}_{[T \times 1]} = \mathbf{V}_{[T \times 1]} \quad (7)$$

where $\varepsilon$ represents the noise at each time step, and is assumed to follow a Gaussian distribution, $\mathcal{N}(0, \sigma^2 \mathbf{I}_{[T*T]})$. An unbiased estimate of $\sigma^2$ can be obtained as follows:

$$\sigma^2 = \frac{(\mathbf{V} - \mathbf{WX})^{\mathrm{T}}(\mathbf{V} - \mathbf{WX})}{T - P} \quad (8)$$

where $T$ and $P$ are the total numbers of time steps and pixels. The difference between these two quantities is the degree of freedom. Combining Equations (6) and (7), $\mathbf{X}$ becomes a Gaussian vector:

$$\mathbf{X} = (\mathbf{W}^\mathrm{T}\mathbf{W} + \lambda\mathbf{I})^{-1}(\mathbf{W}^\mathrm{T}\mathbf{W})\mathbf{U} + (\mathbf{W}^\mathrm{T}\mathbf{W} + \lambda\mathbf{I})^{-1}\mathbf{W}^\mathrm{T}\boldsymbol{\varepsilon} \quad (9)$$

Then, the expectation and covariance matrices of $\mathbf{X}$ can be derived as follows:

$$\mathbf{E}[\mathbf{X}] = (\mathbf{W}^\mathrm{T}\mathbf{W} + \lambda\mathbf{I})^{-1}(\mathbf{W}^\mathrm{T}\mathbf{W})\mathbf{U} \quad (10)$$

$$\mathbf{Cov}[\mathbf{X}] = (\mathbf{X} - \mathbf{E}[\mathbf{X}])(\mathbf{X} - \mathbf{E}[\mathbf{X}])^\mathrm{T} = \sigma^2(\mathbf{W}^\mathrm{T}\mathbf{W} + \lambda\mathbf{I})^{-1}\mathbf{W}^\mathrm{T}\mathbf{W}(\mathbf{W}^\mathrm{T}\mathbf{W} + \lambda\mathbf{I})^{-1} \quad (11)$$

The square root of the diagonal elements in the covariance matrix are the 1-sigma uncertainty values for the retrieved map (Figure S4). The uncertainty map is consistent with the viewing geometry; since the sub-solar and the sub-spacecraft points are always near the equator, pixels at lower latitudes have higher weights, and the uncertainties increase toward the poles. The uncertainty values are on the order of ~10% of the pixel values in the retrieved map (Figure 4a), which suggests a good quality of Earth surface map reconstruction.

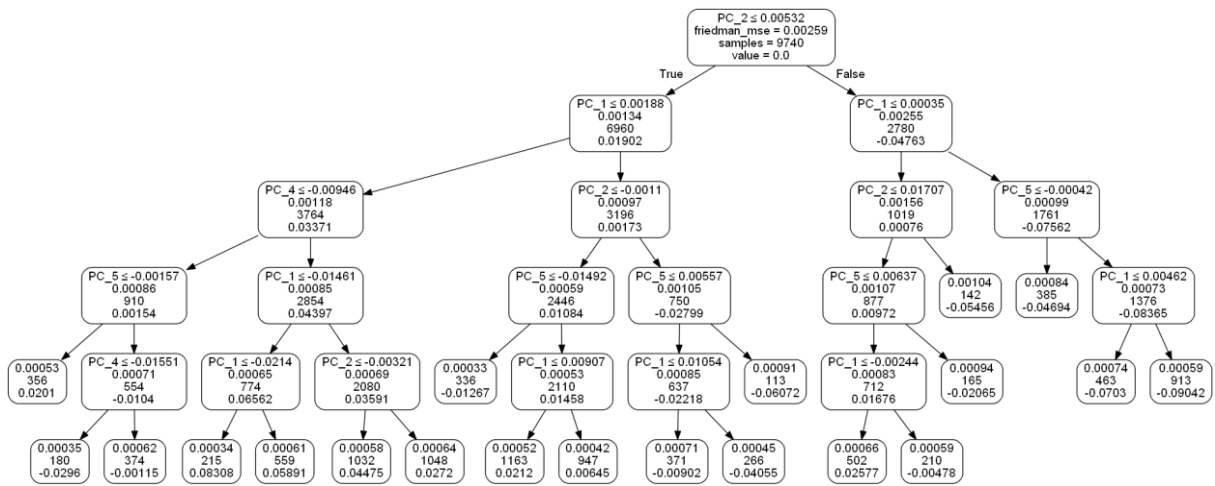

**Figure S1.** The first decision tree in the Gradient Boosted Regression Trees (GBRT) model. The text in each node shows the criterion, the mean square error (MSE), the number of samples and the averaged label value of the node, respectively. The leaf nodes (nodes that do not have child nodes) only have the latter three.

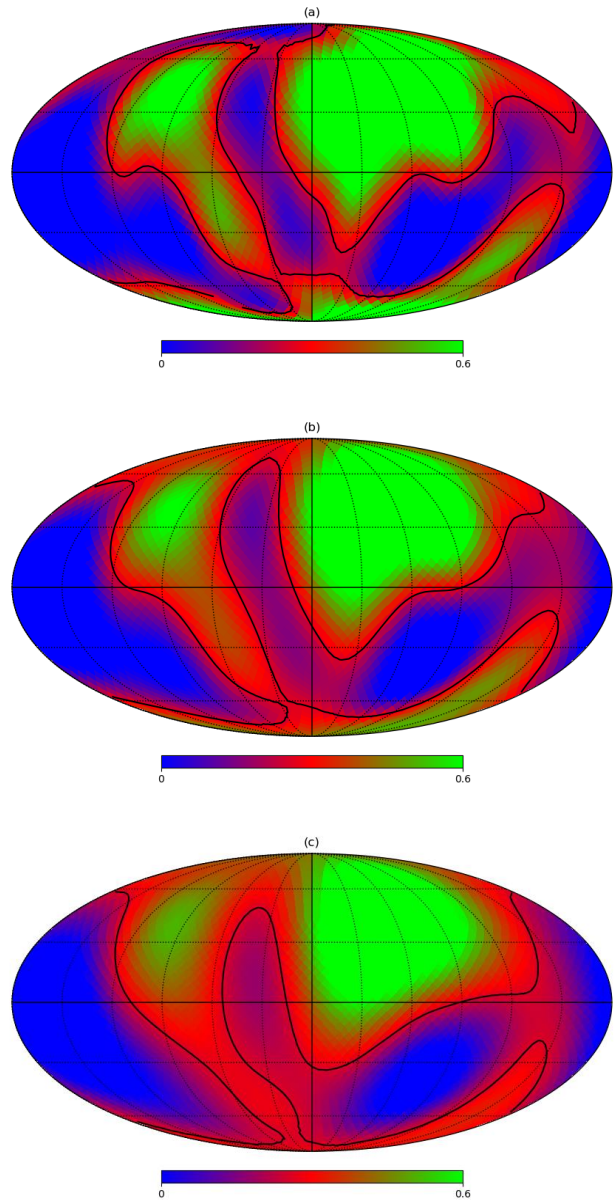

**Figure S2. (a)** Recovered land fraction map using synthetic observations, produced by averaging the ground truth of the land/ocean map given the viewing geometry. The contour of the median value is given by the black line. The regularization parameter, λ, is $10^{-4}$ for constructing this map. (b), (c): Same as (a), but for λ=$10^{-3}$ and $10^{-2}$, respectively.

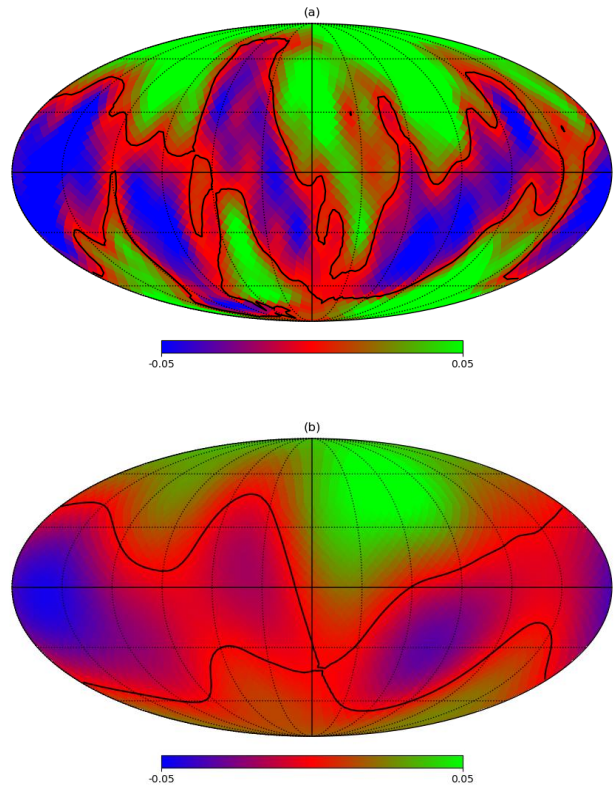

**Figure S3.** (a), (b): Same as Figure 4a, but for $\lambda=10^{-4}$ and $10^{-2}$, respectively.

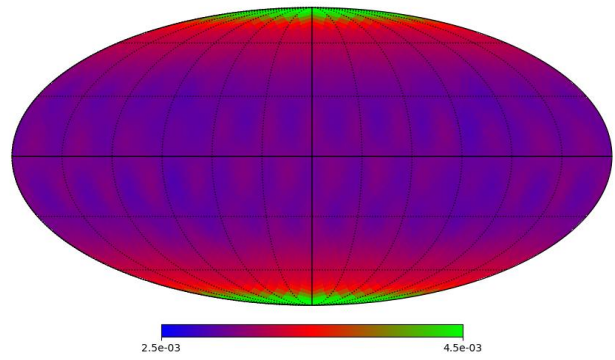

**Figure S4.** Uncertainty in the retrieved Earth surface map shown in Figure 4a.

**References**

Campbell, B., Walker, G. A. H., & Yang, S. 1988, ApJ, 331, 902

Cowan, N. B., Agol, E., Meadows, V. S., et al. 2009, ApJ, 700, 915

Cowan, N. B., & Strait, T. E. 2013, ApJ, 765, L17

Cowan, N. B., Robinson, T., Livengood, T. A., et al. 2011, ApJ, 731, 76

Cowan N.B., & Fujii Y. 2017, Mapping Exoplanets. In: Deeg H., Belmonte J. (eds) Handbook of Exoplanets. Springer, Cham

Farr, B., Farr, W. M., Cowan, N. B., et al. 2018, AJ, 156, 146

Friedman, J. H. 2001, The annals of Statistics, 29, 1189

Fujii, Y. & Kawahara, H. 2012, ApJ, 755, 101

Ge, H., Zhang, X., Fletcher, L. N. et al. 2019, AJ, 157, 89

Geissler, P., Thompson, W. R., Greenberg, R., et al. 1995, JGR, 100, 16895

Gillon, M., Triaud, A. H. M. J., Demory, B.-O., et al. 2017, Nature, 542, 456

Górski, K. M., Hivon, E., Banday, A. J., et al. 2005, ApJ, 622, 759

Jiang, J. H., Zhai, A. J., Herman, J., et al. 2018, AJ, 156, 26

Li, J. -Z., Fan, S., Kopparla, P., et al. 2019, ESS, 6, 166

Lustig-Yaeger, J. Meadows, V. S., Mendoza, G. T., et al. 2018, AJ, 156, 301

Kawahara, H. 2016, ApJ, 822, 112

Knutson, H. A., Charbonneau, D., Allen, L. E., et al. 2007, Nature, 447, 183

Louden, T., & Wheatley, P. J. 2015, ApJ, 814, L24

NASA Exoplanet Archive, Confirmed Planets Table, doi: 10.26133/NEA1

Pedregosa F., Varoquaux, G., Gramfort, A., et al. 2011, JMLR, 12, 2825

Pallé, E., Ford, E. B., Seager, S., Montañés-Rodríguez, P., & Vazquez, M. 2008, AJ, 676, 1319

Platnick, S., King, M., & Hubanks, P. 2015. MODIS Atmosphere L3 Daily Product.

doi:10.5067/MODIS/MOD08_D3.006; doi:10.5067/MODIS/MYD08_D3.006

Sagan, C., Thompson, W. R., Carlson, R., et al. 1993, Nature, 365, 715

Schwartz, J. C., Sekowski, C., Haggard, H. M., et al. 2016, MNRAS, 457, 926

Wessel, P., & Smith, W. H. F. 1996, JGR, 101, 8741